\documentclass{iau} 
\usepackage{graphicx}

\title[IAU 359.~~Red nuggets and globular clusters] 
{Reconstructing the mass accretion histories of nearby Red Nuggets with their globular cluster systems}

\author[Michael A. Beasley et al. ]   
{Michael A. Beasley$^{1,2}$, Ryan Leaman$^3$, Ignacio Trujillo$^{1,2}$,
Mireia Montes$^4$, Alejandro Vazdekis$^{1,2}$, N\'uria Salvador Rusi\~nol$^{1,2}$, Elham Eftekhari$^{1,2}$, Anna Ferr\'e-Mateu$^{5}$
\and Ignacio Martin-Navarro$^{1,2}$}

\affiliation{$^1$Instituto de Astrof\'isica de Canarias, c/ V\'ia L\'actea s/n, E-38250, La Laguna, Tenerife, Spain, \\ email: {\tt beasley@iac.es, trujillo@iac.es, vazdekis@iac.es, nsalva@iac.es, ignacio.martin@iac.es} \\ [\affilskip]
$^2$Departamento de Astrof\'isica, Universidad de La Laguna, E-38205, Tenerife, Spain \\
[\affilskip]
$^3$Max-Planck Institut f\"ur Astronomie, K\"onigstuhl 17, D-69117 Heidelberg, Germany \\ email: {\tt ryan.c.leaman@gmail.com}\\
[\affilskip]
$^4$School of Physics, University of New South Wales, 2052, Sydney, Australia\\ email: {\tt m.montes@unsw.edu.au}\\
[\affilskip]
$^5$Institut de Ci\'encies del Cosmos (ICCUB), Universitat de Barcelona (IEEC-UB), Barcelona 08028, Spain\\
email: {\tt aferremateu@gmail.com}}  
\pubyear{2020}
\volume{359}  
\jname{Galaxy evolution and feedback across different environments}
\editors{T. Storchi-Bergmann, R. Overzier, W. Forman \& R. Riffel, eds.}
\begin{document}

\maketitle

\begin{abstract}
It is generally recognized that massive galaxies form through a combination of {\it in-situ} collapse and {\it ex-situ} accretion. The {\it in-situ} component forms early, where gas collapse and compaction leads to the formation of massive compact systems (blue and red "nuggets") seen at $z>1$. The subsequent accretion of satellites brings in {\it ex-situ}  material, growing  these nuggets in size and mass to appear as the massive early-type galaxies (ETGs) we see locally.  Due to stochasticity in the accretion process, in a few rare cases a red nugget will evolve to the present day having undergone little  {\it ex-situ} mass accretion. The resulting massive, compact and ancient objects have been termed  "relic galaxies". Detailed stellar population and kinematic analyses are required to characterise these systems. However, an additional crucial aspect lies in determining the fraction of {\it ex-situ} mass they have accreted since their formation. Globular cluster systems can be used to constrain this fraction, since the oldest and most metal-poor globular clusters in massive galaxies  are primarily an accreted, {\it ex-situ}  population.  Models for the formation of relic galaxies and their globular cluster systems  suggest that, due to their early compaction and limited accretion of dark-matter dominated satellites, relic galaxies should have characteristically low dark-matter mass fractions compared to ETGs  of the same stellar mass. 
\keywords{galaxies: massive, accretion, evolution, dark matter, globular clusters}
\end{abstract}

\firstsection 
\section{Introduction}

The most massive (log $(M_*/M_\odot) > 11$) galaxies in the nearby Universe are early-type galaxies (ETGs) (e.g. Kauffmann et al. 2003; Kelvin et al. 2014). These  galaxies have large effective radii ($R_e > 5$ kpc) and  generally show extended stellar envelopes (e.g. Caon et al. 1993; Spavone et al. 2017). The inner ($\sim1$ kpc) of ETGs are extremely old,  metal-rich and $\alpha$-element enhanced,  suggestive of a rapid and early dissipational collapse (Trager et al. 2000; Thomas et al. 2003). In addition, the central regions of ETGs appear to be dominated by a dwarf-rich "bottom-heavy" stellar initial mass function (IMF) (Cenarro et al. 2003; Conroy \& van Dokkum 2012; La Barbera et al. 2013) 

In contrast,  mass-matched quiescent galaxies at $z\sim2$ do not appear extended, but are in fact extremely compact systems (Re$\sim1$ kpc)  (Trujillo et al. 2007; van Dokkum et al. 2008;  Damjanov et al. 2009).  These "red nuggets" (Damjanov et al. 2009) show properties remarkably similar to the central regions of ETGs (Trujillo et al. 2014; Mart\'in-Navarro et al. 2015; Toft et al. 2017; Newman et al. 2018). In order to reconcile the high- and low-redshift populations of massive galaxies, red nuggets must grow in effective radius by a factor of $\sim4$ without further significant star formation (Trujillo et al. 2007; van Dokkum et al. 2008). These observations, in conjunction with models of galaxy evolution (e.g., Oser et al. 2010: Shankar et al. 2013; Ceverino et al. 2015) have led to a "two-phase" picture of massive galaxy formation. In the first phase an initial gas collapse, compaction  and quenching event(s) forms a  passively evolving red nugget. In the second phase the red nugget grows in size (and mass) via mergers and satellite accretions to form what we see as a present-day ETG. In this basic picture, the innermost regions of ETGs comprise {\it in-situ} material (stars, gas and dark matter), while the outer envelope is composed of  {\it ex-situ} material originating from accreted satellites (e.g., Mart\'in-Navarro et al. 2019).

\section{Relic galaxies}

The individual accretion histories of massive galaxies are expected to vary quite significantly, and this stochasticity in part drives the scatter seen in $M_* - M_{\rm halo}$ relations at the high mass end of the galaxy mass function (Moster et al. 2010, Behroozi et al. 2013). This stochasticity leads to the prediction that a few, rare objects may essentially skip the second phase of ETG formation entirely such that some red nuggets may reach us from the high redshift universe essentially unaltered (Quilis \& Trujillo 2013). Studying these nearby "relic galaxies"\footnote{Also sometimes called "naked red nuggets" or  massive ultracompact galaxies (MUGs).} will allow us to understand the  stellar populations, kinematics, environments, central black-holes and dark matter content  resulting 
from the first phase of galaxy formation at the highest spatial resolutions and signal to noise.

The first, best sample of nearby relic galaxies was discovered serendipitously in the HETMGS survey (van den Bosch et al. 2015). HETMGS searched for suitable galaxies to measure black hole masses based on sphere-of-influence arguments, which automatically selected for compact  galaxies.
Besides compactness, relic galaxies are observed to be old, metal-rich and have bottom-heavy IMFs at all radii (e.g. Mart\'in-Navarro et al. 2015). Kinematically, they are found to appear as "hot discs" similar to the red nuggets at $z\sim2$ (Toft et al. 2017; Newman et al. 2018) indicative of a predominantly dissipational collapse.
Follow-up observations of the most compact ($R_e < 3$~kpc) and massive (log $(M_*/M_\odot > 11$)  galaxies in HETMGS have produced a sample of 14 galaxies with the above properties (Trujillo et al. 2014; Yildirim et al. 2017; Ferr\'e-Mateu et al. 2017).

Beyond using the structural, chemical and dynamics properties to identify true relic galaxies, ideally one wants to have a handle on their mass accretion histories. By constraining the amount of {\it ex-situ} material in massive galaxies, a clean sample of relic galaxies can be defined. 

\section{Globular clusters in relic galaxies}

\begin{figure}
\begin{center}
 \includegraphics[width=5.5in]{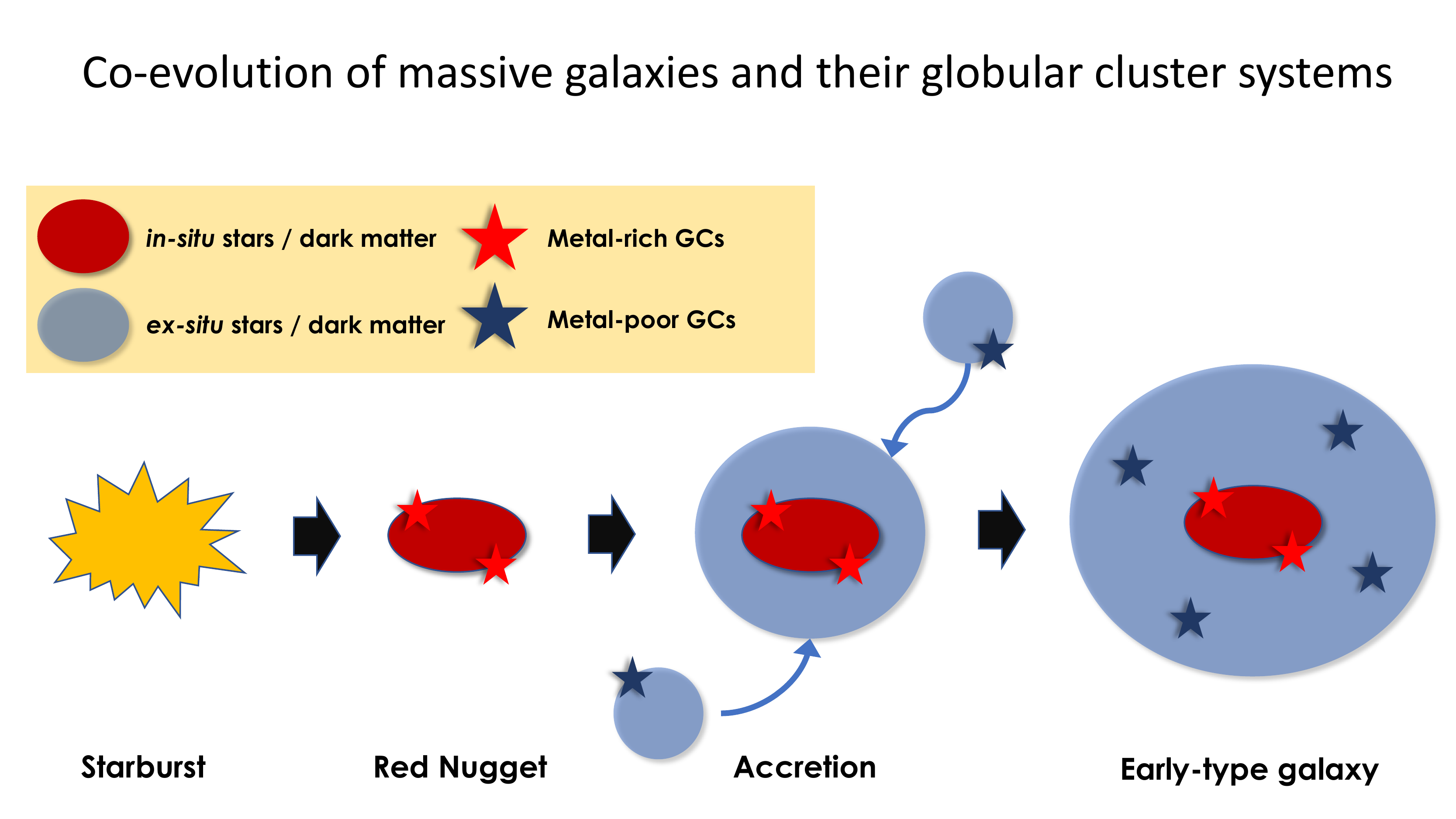} 
 \caption{Cartoon overview of the possible formation and co-evolution of massive galaxies and their globular cluster systems. After an initial gas-compaction, star-burst and subsequent quenching, the resulting "red nugget" is a compact stellar system comprising  {\it in-situ} stars with a retinue of metal-rich globular clusters. Subsequent accretion and merging grows the red nugget in size and mass to form a present-day ETG. The galaxy halo and metal-poor globular clusters come primarily from  accreted {\it ex-situ}  material.}
   \label{fig1}
\end{center}
\end{figure}

Similar to developments in the field of massive galaxy formation, the presently favoured picture for the formation of globular cluster (GC) systems around massive galaxies is also a two phase one. GC systems in massive ETGs show complex colour and metallicity distributions functions (e.g., Peng et al. 2006, Harris et al. 2017). The most metal-rich "red" GCs are predominantly an {\it in-situ} population likely formed during the galaxies' initial collapse. Metal-poor "blue" GCs are found in all but the lowest mass dwarfs, but in massive galaxies are regarded as a largely {\it ex-situ} population brought in via satellite accretion\footnote{The precise mapping between blue and red GCs, and {\it ex-situ} and {\it in-situ} components is more complicated that the picture outlined here (see e.g. Fahrion et al. 2020).}. This general model finds extensive observational and theoretical support (e.g., C\^ot\'e et al. 1998; Beasley et al. 2002; Tonini 2013; Leaman et al. 2013; Mackey et al. 2019; Kruijssen et al. 2019).   A cartoon of the basic scheme is shown in Fig. 1.  

The colour and metallicity distributions of GC systems 
can be readily observed and compared to expectations from hierarchical merger models. This exercise has been performed for the archetypical relic galaxy NGC~1277 in the Perseus galaxy cluster (Beasley et al. 2018). Initially identified as a massive, compact galaxy with an "overmassive" black hole for its stellar mass (van den Bosch et al. 2012), the properties of NGC~1277 have been subsequently shown to be in strikingly good agreement with expectations for a $z\sim2$ red nugget (Trujillo et al. 2014; Mart\'in-Navarro et al. 2015; Yildirim et al. 2017). In terms of its GC system, Beasley et al.~(2018) found that NGC~1277 has very few, if any metal-poor GCs. This observation suggests that the galaxy contains very little  {\it ex-situ} material. By connecting distinct GC subpopulations with {\it in-situ} and {\it ex-situ} origins, a picture of the accreted sub-halo mass function of massive galaxies can be constructed. The colour distributions and total population of GCs in "normal" ETGs suggests accreted mass fractions of 40--70\% (Beasley et al. 2018). This is in broad agreement with other observational approaches  (e.g., Spavone et al. 2017.). By contrast, the inferred accretion fraction for NGC~1277 -- based on its GC system -- is $<12\%$. This is the first relic galaxy for which this technique has been applied, but promises to be a useful approach for identifying true relic galaxies by quantification of their {\it ex-situ} mass fractions. 

\section{Dark matter and the baryon fraction}

An additional, interesting area for investigation concerns the dark matter content of relic galaxies.  As a consequence of their extremely modest {\it ex-situ} fractions, relic galaxies accrete few low stellar mass, dark matter-dominated satellites during their second phase of evolution. Again, this is evidenced by the lack of very metal poor GCs in the GC system of NGC~1277 (Beasley et al. 2018), in addition to a lack of noticeable radial gradients in the IMF of its stellar populations (Martin-Navarro et al.~2015).  Therefore, relic galaxies may be expected to have relatively low dark matter halo mass-to-stellar mass ratios, particularly as one moves out in galactocentric  radius. This has been  suggested by dynamical modelling of these systems with spatially resolved spectroscopic data (Yildirim et al.~2017). Given the above, and the fact that relic galaxies are primarily {\it in-situ} stellar populations and are extremely compact for their stellar mass, these galaxies are ideal places to understand the impact of central super-massive black holes on the efficiency of star formation in massive galaxies.

\section{Conclusions}

The massive, compact galaxy NGC~1277 has been identified as the first true "relic" galaxy in the nearby Universe. It presents itself as a direct counterpart to the red nuggets seen at $z\sim2$. A combination of detailed structural,  stellar population and kinematic analyses has been required to properly characterize this galaxy (Trujillo et al. 2014;  Martin-Navarro et al. 2015; Yildirim et al. 2017), while its GC system has proved key to constrain its accretion history (Beasley et al.~2018).  A detailed exploration of the X-ray properties of NGC~1277 (c.f.,  Buote \& Barth 2019) may bring important constraints on its dark matter halo, as will dynamical studies at large radii via kinematics of the GC system with {\it James Webb Space Telescope}.  Upcoming analysis of its stellar populations in the UV will be used to search for possible star formation even at the sub- one percent level (e.g., Salvador-Rusi\~nol et al. 2019), and infrared spectroscopy will bring further insights into the properties of its stellar populations. 

The presence of NGC~1277 in the Perseus cluster places a lower limit on the space densities of relic galaxies, $\rho_{\rm relic}>7\times10^{-7}$ Mpc$^{-3}$,  and we note that several more relics have now been identified (Ferr\'e-Mateu et al. 2017; Yildirim et al. 2017). A study of the GC populations of these galaxies will be crucial for constraining their accretion fractions. The identification and study of relic galaxies provide the opportunity to study a population normally only accessible at $z>2$, and  gain a detailed understanding of the earliest phases of massive galaxy formation.

\end{document}